\documentclass[12pt, draftclsnofoot, onecolumn]{IEEEtran}
\usepackage{cite}
\usepackage{amsmath,amssymb,amsfonts}
\usepackage{algorithmic}
\usepackage{graphicx}
\usepackage{tabularx}
\usepackage{booktabs}
\usepackage{textcomp}
\usepackage{todonotes}
\usepackage{paralist}
\usepackage{xcolor}
\usepackage{xfrac}
\usepackage[caption=false]{subfig}
\begin{document}
\title{RF-Assisted Free-Space Optics for\\ 5G Vehicle-to-Vehicle Communications}

\author{\IEEEauthorblockN{Mattia~Brambilla\IEEEauthorrefmark{1}, Andrea~Matera\IEEEauthorrefmark{1}, Dario~Tagliaferri\IEEEauthorrefmark{1}\IEEEauthorrefmark{2}, Monica~Nicoli\IEEEauthorrefmark{3}~and~Umberto~Spagnolini\IEEEauthorrefmark{1}}  \\
	 \IEEEauthorblockA{\IEEEauthorrefmark{1}Dipartimento di Elettronica, Informazione e Bioingegneria (DEIB), Politecnico di Milano, Italy}
	\IEEEauthorblockA{\IEEEauthorrefmark{2}Istituto di Elettronica e di Ingegneria dell'Informazione e delle Telecomunicazioni (IEIIT), CNR,  Italy}
	\IEEEauthorblockA{\IEEEauthorrefmark{3}Dipartimento di Ingegneria Gestionale (DIG),  Politecnico di Milano, Italy} \\
		Email: \{mattia.brambilla,andrea.matera,monica.nicoli,umberto.spagnolini\}@polimi.it, dario.tagliaferri@ieiit.cnr.it}
	
\maketitle
\begin{abstract}
	Vehicle-to-Vehicle (V2V) communications are being proposed, tested
	and deployed to improve road safety and traffic efficiency. However,
	the automotive industry poses strict requirements for safety-critical
	applications, that call for reliable, low latency and high data rate
	communications. In this context, it is widely agreed that both Radio-Frequency
	(RF) technologies at mmWaves and Free-Space Optics (FSO) represent
	promising solutions, although their performances are severely degraded
	by transmitter-receiver misalignment due to the challenging high-mobility
	conditions. By combining RF and FSO technologies, this paper proposes
	a FSO-based V2V communication system where the pointing coordinates of laser sources
	are based on vehicle's information exchanged over a reliable low-rate
	RF link. Numerical simulations demonstrate that such
	compensation mechanism is mandatory to counteract the unavoidable
	misalignments induced by vehicle dynamics, and thus to enable FSO
	technology for V2V communications even in high mobility scenarios. 
\end{abstract}


\section{Introduction}

Vehicular communications are nowadays gaining wide attention from
both industry and academia, as they represent a leading market for
the fifth generation (5G) of cellular systems. Cooperative Intelligent Transportation Systems (C-ITS) aim at improving road safety and traffic efficiency by Vehicle-to-everything (V2X) communications, which enable fast sharing of massive mobility data through vehicular cloud networks \cite{Festag2015,Brambilla2018}.
Differently from any other type of wireless
systems, V2X, and in particular Vehicle-to-Vehicle (V2V)
communications, are extremely critical due to the very high
mobility of terminals and to the strict requirements  in terms of end-to-end latency, reliability and data rate needed to support 5G C-ITS services \cite{3GPP_Rel16_2}.
A possible solution to the challenging demands of V2V Ultra Reliable
Low Latency Communications (URLLC) is to exploit the millimeter Wave
(mmWave) band \cite{Heath2016}. However, high-frequency links are
subject to severe path loss which severely limits the communication range. To face it, Massive Multiple-Input Multiple-Output (MIMO)
antenna architectures along with sophisticated beamforming techniques have been considered to support mmWave V2V communications \cite{Perfecto2017}.
Optical Wireless Communications (OWC), and in particular laser-based
Free-Space Optics (FSO), are emerging as a promising alternative/complementary V2V
solution. FSO is in fact a mature technology
offering a huge bandwidth (over an unlicensed spectrum) with relatively
low cost, small size and ease of deployment, by using simple Intensity-Modulation/Direct-Detection
(IM/DD) transmission schemes, such as On-Off Keying (OOK), and,
lastly, it does not suffer from Doppler shift as highly-dynamic RF communications systems \cite{Kaushal-2017}. Furthermore, the
extremely high directivity of laser sources allows to establish dense
networks of high-capacity, high-security links with minimized mutual interference and eavesdropping.  As such, FSO is emerging as an attractive candidate to interconnect fixed base stations by point-to-point links in next-generation cellular networks, as well as dynamic devices such as drones \cite{Kaadan-2014}, vehicles \cite{Epple-2007} and airplanes  \cite{F.Moll-LaserCommAircraft-2015}. 
\begin{figure}[!t]
	\centering \includegraphics[width=0.7\linewidth]{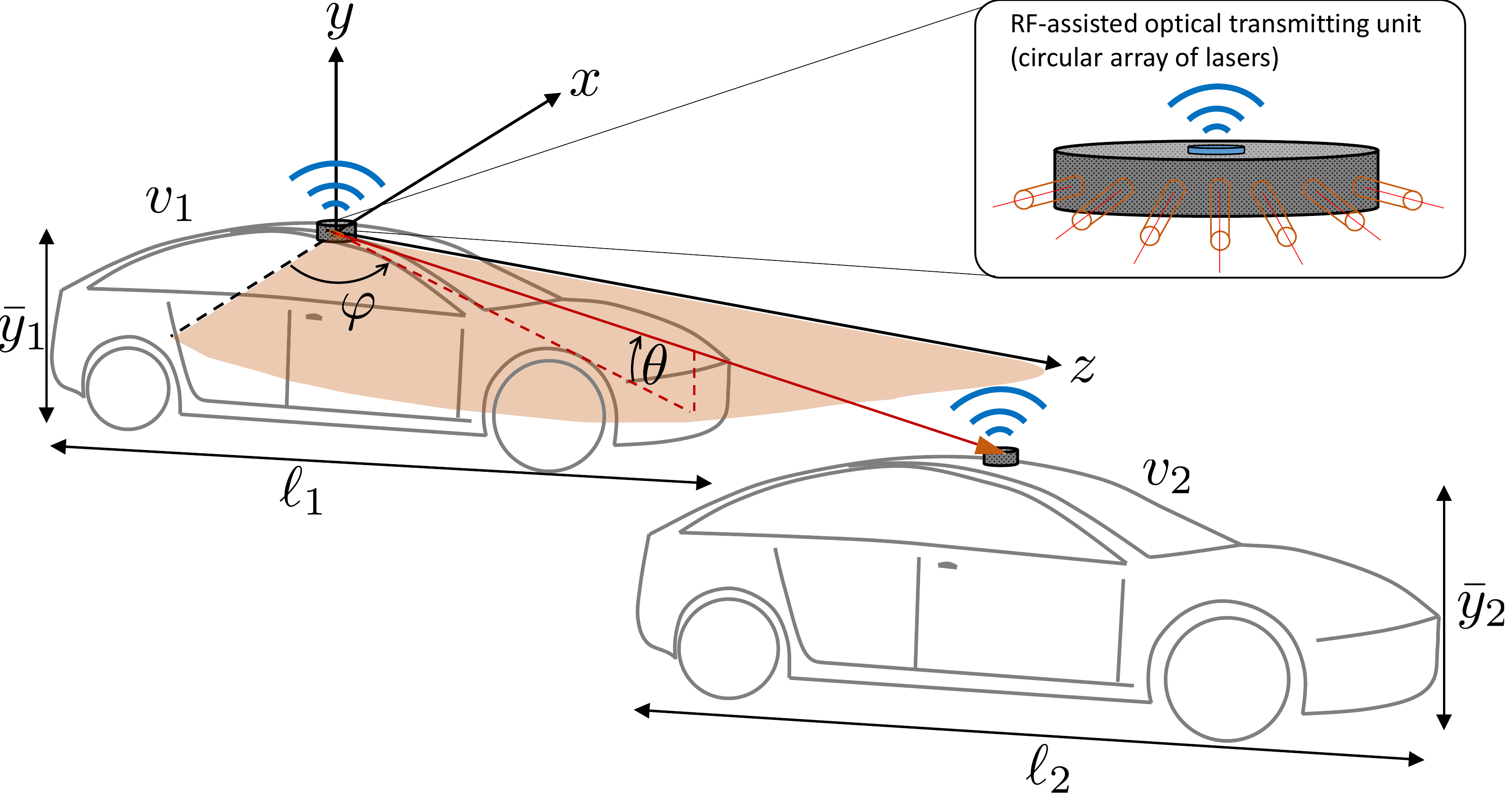}
	\caption{Geometry of the V2V FSO communication system. To achieve the requested
		angular span in the horizontal direction, a circular array of laser
		is employed.}	
	\label{fig:vehiclegeometry} 
\end{figure}

The
deployment of FSO links in V2V scenarios poses two main challenges:
\textit{i)} FSO links among multiple moving targets are highly hindered
by Tx-Rx misalignments due to the very narrow beams, and this should be
properly taken into account in the system design, and \textit{ii)}
adverse atmospheric conditions, especially fog, mist and haze, severely
degrade the performance as the propagating beam may undergo to absorption
and scattering phenomena, resulting in deep signal fading \cite{Esmail-2016}.
While latter effects are hardly avoidable, Tx-Rx misalignment in FSO
V2V communications can be faced either by widening the beam to enhance the coverage area at the receiver, or by advanced Active Tracking
and Pointing (ATP) mechanisms. Beam enlargement has the drawback of reducing the power density and increasing interference, which lead to performance degradation if not properly handled \cite{Tagliaferri-Matera-2018}. In contrast, ATP is able to guarantee the continuous Line-Of-Sight (LOS) condition while exploiting the inherent laser directivity, and therefore the paper focuses on this solution. Among the numerous
ATP techniques recently proposed \cite{Kaymak-2018}, the one based on
Micro Electro-Mechanical Systems (MEMS) micromirrors is particularly
attractive as it guarantees small size, low weight, accuracy, low
power consumption, position repeatability and relatively fast response
(up to tens of KHz)\cite{Mirrorcle-2018}. Based on the information
exchanged among vehicles over a dedicated low-rate control RF link,
MEMS-based mirrors are considered for the proposed FSO-based V2V communication
scheme to compensate for small-to-medium angular pointing errors (up to $\pm10$
degrees).

\paragraph*{Contributions}

The contribution of this paper is three-fold: \textit{i)} we propose
a FSO-based V2V communication system where the pointing of laser sources
is carried out by exploiting side information about vehicles'
positions exchanged among vehicles over a
low-rate RF-link, \textit{ii)} we analyze the robustness of a V2V
FSO system with respect to Tx-Rx misalignments, concluding that  reliable alignment in high mobility conditions is hardly achievable
without the help of the proposed pre-compensation mechanism based
on the information retrieved by the RF link, and \textit{iii)}
we confirm the benefits of the proposed hybrid FSO-RF system
by numerical simulations of realistic V2V links  between vehicles in a platoon formation.

\paragraph*{Organization}
The paper is organized as follows. Sec. \ref{sec:geometrical settings} and \ref{Sec: Optical channel model} introduce the modeling of vehicle dynamics and the FSO channel, respectively.
Sec. \ref{sec:case studies} details different V2V communication strategies to compensate misalignments. Lastly, Sec. \ref{sect:results} presents the numerical results.

\section{System model}\label{sec:geometrical settings}

As illustrated in Fig. \ref{fig:vehiclegeometry}, we consider two vehicles, $v_1$ and $v_2$, of length $\ell_1$ and $\ell_2$ and height $\bar{y}_1$ and $\bar{y}_2$, respectively. Tx and Rx FSO units are assumed to be installed on top of vehicles $v_1$ and $v_2$, respectively.

We adopt a $v_1$-based, right-handed coordinate system with the origin located in the center of the Tx unit and solid with the laser, propagating along the $z$-axis. This coordinate system is particularly useful to handle high-mobility scenarios, where the laser can be dynamically steered.  
In order to evaluate the performance of the V2V FSO communication system, the azimuth and elevation angles with respect to the chosen coordinate system have to be defined as functions of the relative position of the vehicles. The azimuth angle $\varphi $ is defined on the $zx$-plane, while the elevation angle $\theta$ is referred to the $zy$-plane. Defining the 3D location of vehicle $v$ by the position of its FSO transceiver unit, denoted as ($x_v, y_v, z_v$), the azimuth and elevation angles at time $t$ are given by:
\begin{align}
\varphi [t] & =\text{tan}^{-1}\left(\frac{x_{2}[t]-x_{1}[t]}{|z_2[t]-z_1[t]|}\right) =\text{tan}^{-1}\left(\frac{x_{2}[t]-x_{1}[t]}{z[t]}\right),
\end{align}
\begin{align}
\theta[t] & =\text{tan}^{-1}\left(\frac{y_{2}[t]-y_{1}[t]}{|z_2[t]-z_1[t]|}\right) =\text{tan}^{-1}\left(\frac{y_{2}[t]-y_{1}[t]}{z[t]}\right),
\end{align}
where $z[t] = |z_2[t]-z_1[t]|$ well approximates the V2V distance, as commonly assumed when treating laser beam propagation \cite{Simon-2016}. 
In order to let the FSO system to work properly, it is essential that the transmitter knows the values $\varphi [t]$ and $\theta[t]$, for each time instant $t$. We assume that the two vehicles exchange over a dedicated RF service channel the information on their mutual position in order to tune the pointing directions. To this purpose, we consider to employ a MEMS micromirror, which allows to steer a laser within a medium-to-low angular range (typically, $\pm 10$\textdegree) in both horizontal and vertical directions. Since, in general, horizontal variations of the pointing are much larger than the vertical ones due to the relative drifting of the two vehicles during their motion, we assume the presence of multiple laser beams placed in a circular-like array as in Fig. \ref{fig:vehiclegeometry}. This lasers configuration, in fact, allows to have sufficient angle span in the azimuth direction. The sizing of the angular spacing between two adjacent laser elements is assumed to enable a continuous sampling of the horizontal direction, accounting for both the beam divergence and the MEMS' steering capability; a design is beyond the scope of this paper.

\begin{figure}[!tb]
	\centering
	\includegraphics[width=0.7\linewidth, keepaspectratio]{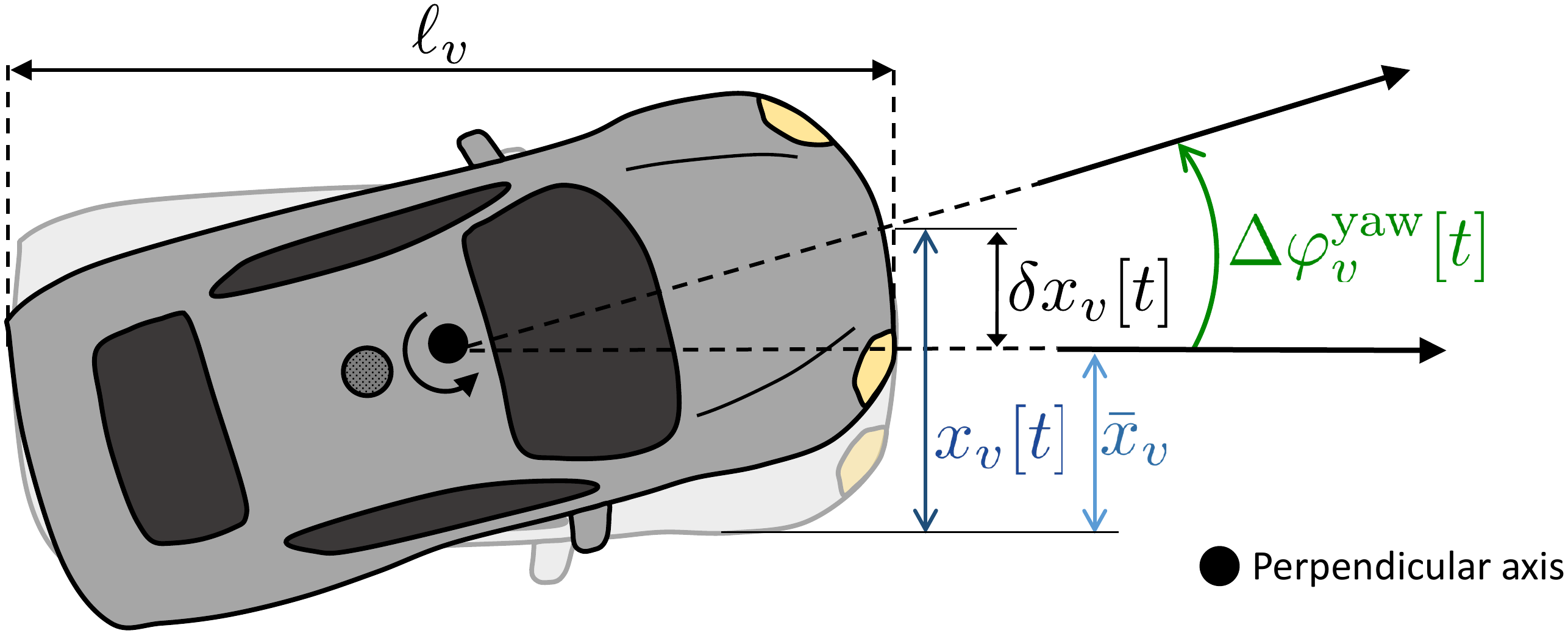}
	\caption{Vehicle yaw as a rotation around the perpendicular axis.}
	\label{fig:Vehicle_Yaw}
\end{figure}

\begin{figure}[!tb]
	\centering
	\includegraphics[width=0.7\linewidth, keepaspectratio]{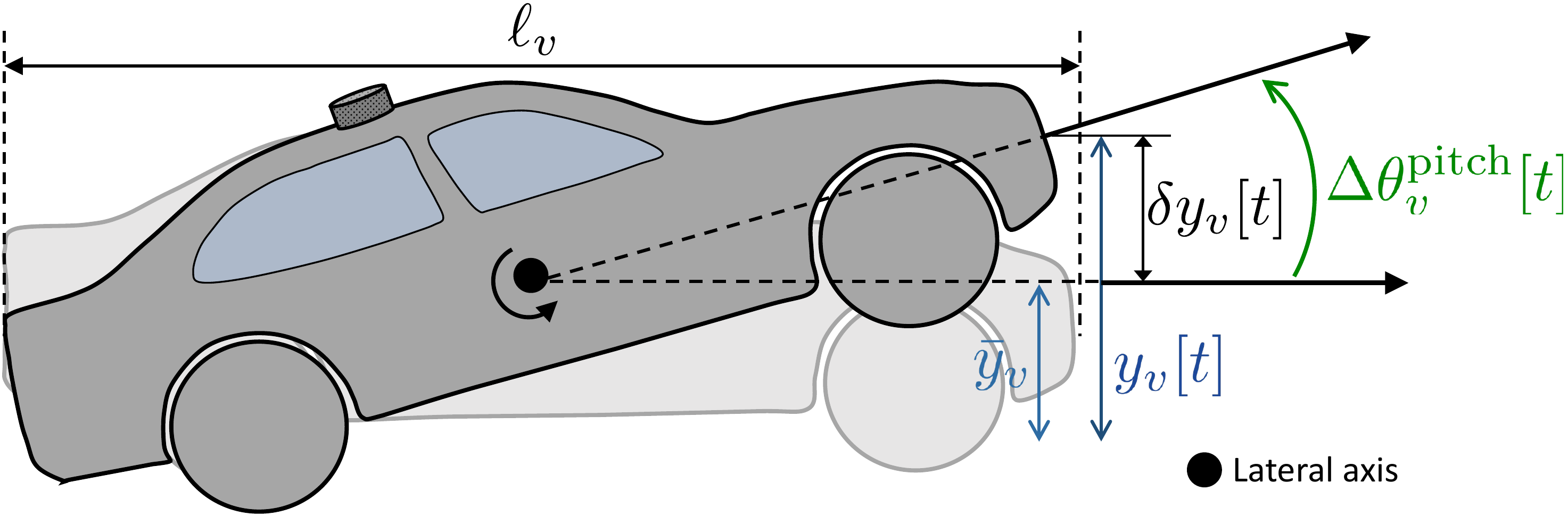}
	\caption{Vehicle pitch as a rotation around the lateral axis.}
	\label{fig:Vehicle_Pitch}	
\end{figure}

Once identified the couple $(\varphi,\theta)$, only the laser that better matches the selected azimuth angle is kept active for each vehicles pair, and its elevation is modified accordingly.
After this selection phase, a tracking phase takes place, where the laser orientation is dynamically updated using the RF link in order to compensate the vibrations and tilting of vehicles due to, for example, engine vibrations, road conditions, electro-mechanical vehicle configurations and so forth. If not properly compensated, these perturbations impact on the laser pointing angles $(\varphi,\theta)$ and, as a consequence, on the efficiency of the FSO system, as shown later on in this paper. In particular, any fluctuations $\delta x_v [t]$ on the horizontal plane causes the vehicle yaw, \textit{i.e.,} a rotation along the lateral axis of the vehicle ($x$-axis),
and it impacts on the azimuth angle $\varphi$. Similarly, any perturbation along the $y$-axis causes a rotation around the lateral axis ($y$-axis) affecting the elevation angle $\theta$. 
These variations are modeled as fluctuations in time ($\delta x_v [t]$ and $\delta y_v [t]$)  over a mean value  ($\bar{x}_v$ and $\bar{y}_v)$ corresponding to the condition at rest, and they are given by:
\begin{align}
x_v [t] = \bar{x}_v + \delta x_v [t] , \qquad
y_v [t] = \bar{y}_v + \delta y_v [t].
\label{eq: x-y}
\end{align}
Figs. \ref{fig:Vehicle_Yaw} and \ref{fig:Vehicle_Pitch} illustrate these fluctuations as a consequence of the perturbation at rest. In particular, Fig. \ref{fig:Vehicle_Yaw} focuses on the vehicle yaw $\Delta\varphi_{v}^{\mathrm{yaw}}[t]$ while Fig. \ref{fig:Vehicle_Pitch} illustrates the vehicle pitch $\Delta\theta_{v}^{\mathrm{pitch}}[t]$.
The yaw and pitch angles are evaluated, in first approximation, as:
\begin{align}
\Delta\varphi_{v}^{\mathrm{yaw}}[t]=\text{tan}^{-1}\bigg(\frac{x_{v}[t]-\bar{x}_{v}}{0.5\hspace{0.1cm}\ell_{v}}\bigg)= \text{tan}^{-1}\bigg(\frac{\delta x_v [t]}{0.5\hspace{0.1cm}\ell_{v}}\bigg)
\label{eq:yaw angle},
\end{align}
\begin{align}
\Delta\theta_{v}^{\mathrm{pitch}}[t]=\text{tan}^{-1}\bigg(\frac{y_{v}[t]-\bar{y}_{v}}{0.5\hspace{0.1cm}\ell_{v}}\bigg) = \text{tan}^{-1}\bigg(\frac{\delta y_v[t]}{0.5\hspace{0.1cm}\ell_{v}}\bigg).
\label{eq:pitch angle}
\end{align}

\begin{figure}[!tb]
	\centering
	\includegraphics[width=0.7\linewidth, keepaspectratio]{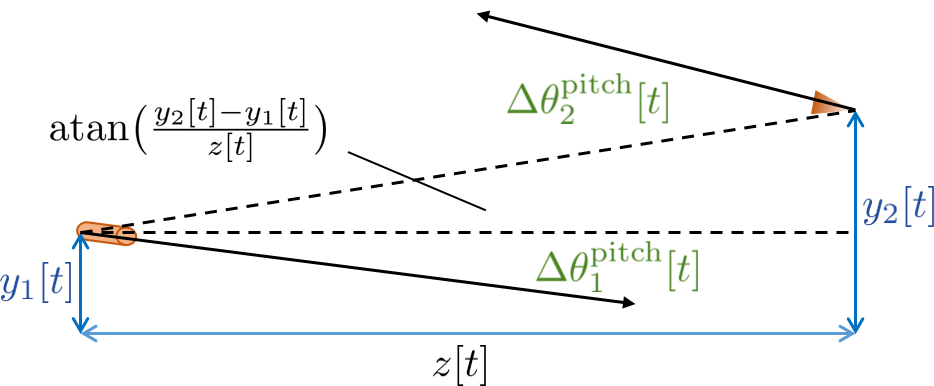}
	\caption{Geometry on the vertical plane with emphasis on the pitch angles.}	
	\label{fig:beta}
\end{figure}

Besides affecting the orientation of laser pointing at vehicle $v_1$, the perturbations in  \eqref{eq:yaw angle}-\eqref{eq:pitch angle} have an unavoidable geometrical impact on the efficiency of the FSO link. In fact, in order to maximize the received power, the light should impinge the receiver orthogonally to its surface, but the perturbations in \eqref{eq: x-y} reduce the effective area of the photodetector. Their effects are taken into account by projecting the receiver's surface in the laser propagation direction. Accordingly, we introduce two angles, namely $\beta_x$ and $\beta_y$, that describe the rotation of the receiver around its local coordinate system:
\begin{align}
\beta_y [t] = & \big|  \Delta\theta_{2}^{\mathrm{pitch}}[t] - \Delta\theta_{1}^{\mathrm{pitch}}[t] \big| ,
\label{beta_y}
\end{align}
\begin{align}
\beta_x [t] = & \big| \Delta\varphi_{2}^{\mathrm{yaw}}[t] - \Delta\varphi_{1}^{\mathrm{yaw}}[t] \big|.
\label{beta_x}
\end{align}
The impact of the  geometry of the Tx-Rx arrangement  as a consequence of the relative vehicle position as well as  the rotations due to vibrations and tilting along the $y$-axis is illustrated in Fig. \ref{fig:beta}. Similar analysis applies  to the $x$-axis.

\section{Optical Channel Model}
\label{Sec: Optical channel model}

\begin{figure}[!t]
	\centering
	\subfloat[Perspective view]{\includegraphics[width=0.7 \linewidth]{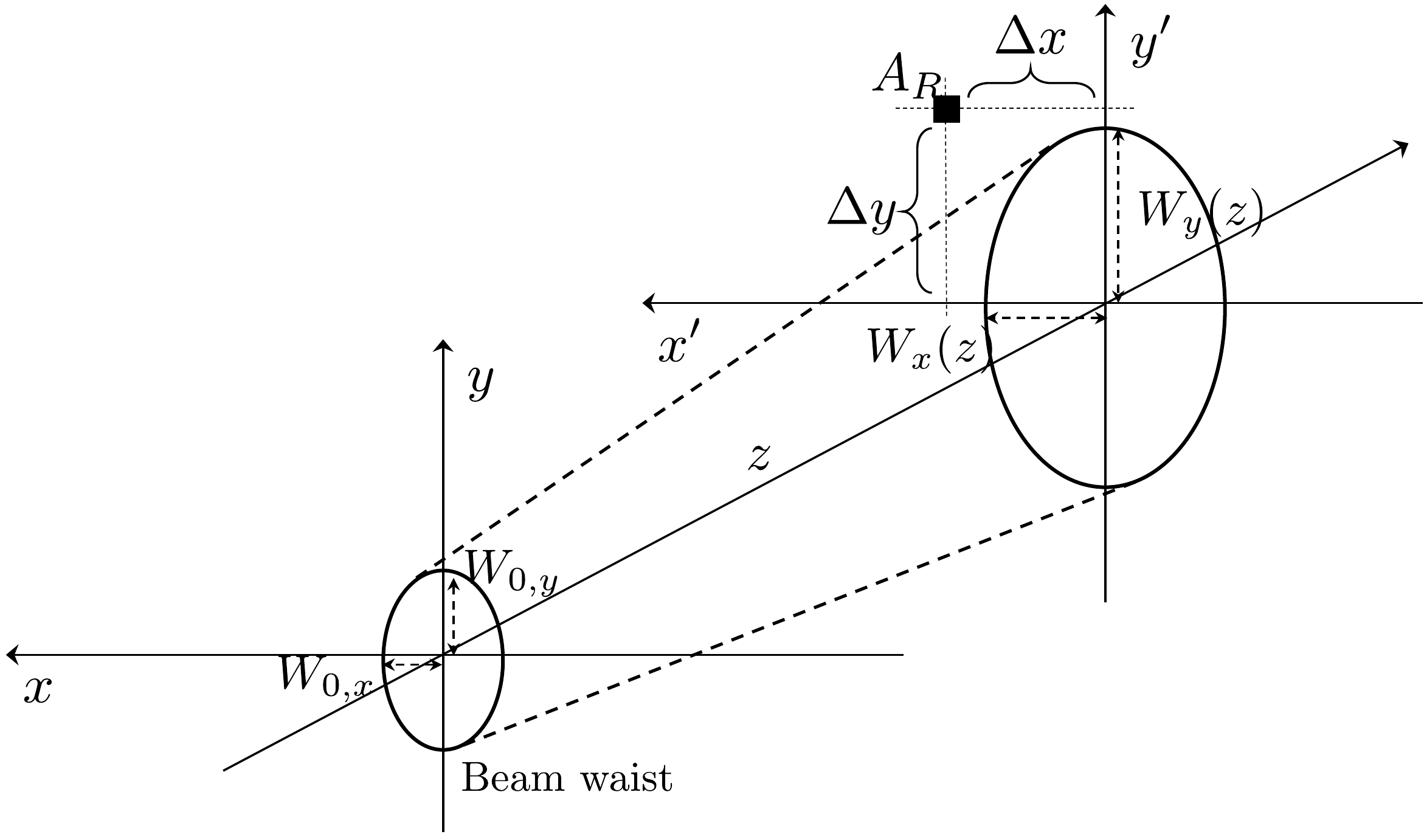}\label{subfig:perspective_gauss}}\\ 	
\subfloat[Lateral view]{\includegraphics[width=0.7\linewidth]{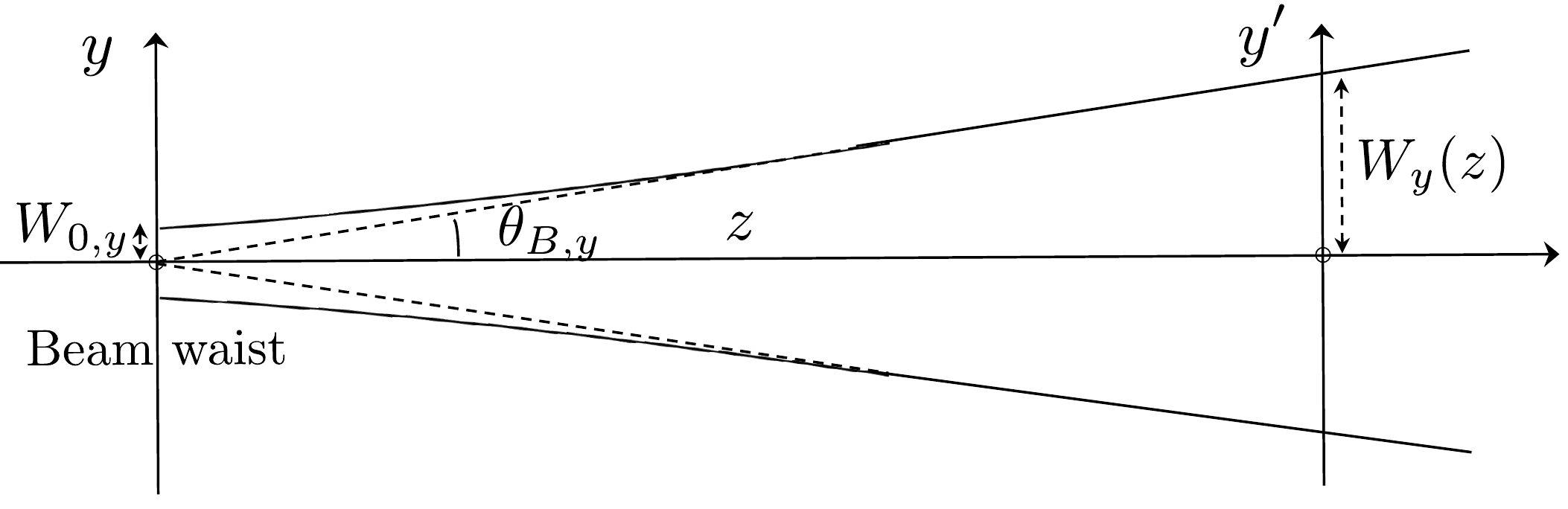}\label{subfig:lateral_gauss}}\\
	\caption{Illustrative sketch of the geometry of a Gaussian propagating beam.}	
	\label{fig:gaussian}
\end{figure}

As first assessment of the link budget, we consider a FSO transmission operating in clear sky conditions (neither turbidity nor turbulence). The signal emitted by a single-mode laser working at wavelength $\lambda$ and emitting a power $P_0$, can be well approximated by a Gaussian beam (first Transverse Electromagnetic mode, TEM$_{00}$), whose propagation along the reference direction $z$ is described by diffraction theory. 
The divergence angles along the transversal $x$ and $y$ coordinates are equal to:
\begin{align}
\theta_{B_x} = \frac{\lambda}{\pi W_{0_x}}, \hspace{0.4cm}
\theta_{B_y} = \frac{\lambda}{\pi W_{0_y}}  ,
\end{align}
where $W_{0_x}$ and $W_{0_y}$ are the beam waist sizes (minimum widths) of the beam along $x$ and $y$, respectively, defining an ellipse on the plane normal to the laser direction enclosing 86.5\% of $P_0$. Assuming the laser beam waist in $z=0$ (at the transmitter), the expression of the light intensity at a distance $z$ is:
\begin{align}
I(z,x,y) = \frac{2 P_0}{\pi W_x(z) W_y(z) } \hspace{0.1cm}\mathrm{e}^{-\frac{2 x^2}{W_x(z)}}\hspace{0.1cm}\mathrm{e}^{-\frac{2 x^2}{W_y(z)}} ,
\end{align}
in which the transversal power decay is controlled by the spot size parameters:
\begin{align}
W_x(z) = W_{0_x} \, \sqrt{1 + \left(\frac{\lambda z}{\pi W_{0_x}^2}\right)^2} \approx \theta_{B_x} z    ,
\end{align}  
\begin{align}
W_y(z) = W_{0_y} \, \sqrt{1 + \left(\frac{\lambda z}{\pi W_{0_y}^2}\right)^2} \approx \theta_{B_y} z  ,
\end{align}
which are linearly dependent on $z$ if $z \gg \pi W_{0,x/y}^2 / \lambda$ (far-field condition). The signal is captured by the receiver, whose effective area $A_R$ is set by the outer lens of the telescope (in general, the set of lenses) in charge of focusing the incoming light on a photodetector having an active area equal to $A_{PD}$. The received power is function of  the receiver displacements from the laser propagation direction (ideal pointing) and its local rotation angles ($\beta_x$ and $\beta_y$). The displacements along $x$ and $y$ are described by the quantities $\Delta x$ and $\Delta y$ (Fig. \ref{fig:gaussian}), that can be computed as  (paraxial approximation):
\begin{align}\label{eq:Delta x}
\Delta x [t] = x_2 [t] - x_1[t] - \tan \big( \Delta\varphi_{1}^{\mathrm{yaw}}[t]\big) z[t],
\end{align} 
\begin{align}\label{eq:Delta y}
\Delta y [t] = y_2 [t] - y_1[t] - \tan \big( \Delta\theta_{1}^{\mathrm{pitch}}[t]\big) z[t]. 
\end{align}
These parameters are essential as they can limit the V2V FSO performance: the road conditions, electro-mechanical vehicle configurations, engine vibrations and so forth impair the LOS communication and result in strong fluctuations of the received optical power. The amount of collected power is calculated as the integral over the effective area of the receiver (projection along the $z$ axis of $A_R$, here expressed as $A_R = L_{R,x}L_{R,y}$ for analytical convenience). Neglecting  the time dependence for the sake of simplicity, the received power $P_R$ is:
\begin{equation}\label{eq:received_power}
\begin{split}
P_R(z,\Delta x,\Delta y) & = \cos\beta_x \cos\beta_y  \cdot \\
& \cdot \int_{\Delta x -\frac{L_{R,x}}{2} }^{\Delta x +\frac{L_{R,x}}{2} } \int_{\Delta y -\frac{L_{R,y}}{2} }^{\Delta y +\frac{L_{R,y}}{2} } I (z,x',y') \,\mathrm{d}x' \mathrm{d}y' \approx \\
&\approx I (z,\Delta x,\Delta y) A_R \cos\beta_x \cos\beta_y  ,
\end{split}
\end{equation}
where the last approximation holds for the receiver collection area  much smaller than the laser spot, \textit{i.e.} $A_R \ll W_x(z)W_y(z)$. 
The performance of the system are evaluated in terms of the Signal-to-Noise Ratio (SNR) after the photodetector for a Positive-Intrinsic-Negative (PIN)-based receiver. The SNR is expressed as:
\begin{equation}
\mathrm{SNR} = \frac{P_R^2 \, \eta^2 }{2\,e\,B \left(\underbrace{E_{b}\,\Delta\lambda}_{
		I_{b}}\,A_R\, \eta + P_R\, \eta\right) + \mathrm{NEP}^2 \, \eta^2 \, B},
\end{equation}
where:
\begin{itemize}[\hspace{-0.05cm}\small $\bullet$]
	\item the numerator is the squared electrical current produced by the signal incident on the photodetector, the latter having a responsivity $\eta$ as function of $\lambda$;
	
	\item the first term at the denominator is the shot noise associated to the background light-induced current (\textit{i.e.,} the sum of the direct and the diffuse solar radiation), and the useful signal. Symbols $e$ and $B$ denote, respectively, the electron's charge and the communication bandwidth. The solar irradiance $I_{b}$ $[\mathrm{W/m^2}]$ is obtained by multiplying the spectrum $E_{b}$ $[\mathrm{W/m^2/nm}]$ and the receiver's optical bandwidth  $\Delta\lambda$ (limited by the responsivity or by a proper optical filter);
	
	\item the second term at the denominator is the current noise power comprising both the dark current of the photodetector and the overall electronic noise generated by the receiving circuitry (mostly from the first amplifying stage). It is summarized by the Noise Equivalent Power $\mathrm{NEP}$ $[\mathrm{W/\sqrt{Hz}}]$.
	
\end{itemize}

In order to evaluate the solar radiation level at the receiver, we made use the Gueymard physical model \cite{Gueymard-2001}, which takes into account the propagation of the sunlight through the atmosphere, and predicts the average amount of absorption and scattering by atmospheric constituents and components, as function of the geographical location, period of the year and hour of the day. As output, we obtain the direct and diffuse solar spectra on a generically tilted surface at a certain height above sea level. Once obtained the SNR, the corresponding throughput $R$ $[\mathrm{bit/s}]$ is computed from Shannon capacity:
\begin{equation}
R = 0.5 \,  B \hspace{0.1cm} \text{log}_2 \big( 1+ \text{SNR} \big).
\end{equation}

\section{Case Studies}
\label{sec:case studies}
The proposed V2V FSO system is analyzed in three conditions characterized by different capabilities of compensating misalignments. We focus on the communication between two vehicles traveling straight ahead in a platoon formation, where $v_2$ is the leader and $v_1$ is the follower. 
A detailed description of the considered cases is provided hereafter.

\subsection{Case 1: V2V FSO with no pointing compensation}\label{subs:case1}

This case represents a condition where vehicle $v_1$ is unable to orientate the laser to the desired direction: the laser is solid with the vehicle chassis and it always points at a frontal direction. 
Clearly, this is the worst possible case, especially in high mobility applications.
In this condition, the displacements $\Delta x [t]$ and $\Delta y [t]$ are given by \eqref{eq:Delta x}$-$\eqref{eq:Delta y}.

\subsection{Case 2: V2V FSO with static compensation}\label{subs:case2}

This case refers to a scenario where a compensation of the nominal values $\bar{y}_v$ and $\bar{x}_v$ is used. To this purpose, the RF link is expected to provide information on the vehicle parameters ($\bar{x}_{1}, \bar{y}_{1}, \bar{x}_{2}, \bar{y}_{2}$) before the beginning of FSO transmission. The perturbations $\delta x_v [t]$ and $\delta y_v [t]$ are the only unknown parameters which can not be compensated. Thus, this case depicts a common condition in mobility where the transmitter knows the location of the receiver but the presence of undesired and unpredictable perturbations impairs the performance. Recalling \eqref{eq: x-y}, by compensating the nominal values $\bar{x}_{v}, \bar{y}_{v}$, the displacements $\Delta x[t]$ and $\Delta y[t]$ become:
\begin{align}
\Delta x[t] = \delta x_2 [t] - \delta x_1 [t]- \mathrm{tan} \big( \Delta\varphi_{1}^{\mathrm{yaw}}[t]\big) z[t]  ,
\end{align} 
\begin{align}
\Delta y [t] = \delta y_2 [t] -\delta y_1 [t] - \mathrm{tan} \big( \Delta\theta_{1}^{\mathrm{pitch}}[t]\big) z[t].
\end{align}

\subsection{Case 3: RF-assisted V2V FSO with dynamic compensation}\label{subs:case3}

This case study assumes a RF channel working in parallel to the FSO link in which vehicles exchange information of the instantaneous location along each axis. In this regard, this case study assumes the availability of on-board sensors to measure the vehicle dynamics. By using sensors' information, $v_1$ is able to compensate its own vibrations and tilting: namely, $\delta x_1 [t] =0$, $ \delta y_1 [t] = 0$, $\Delta\theta_{1}^{\mathrm{pitch}}[t]=0$ and $\Delta\varphi_{1}^{\mathrm{yaw}}[t]=0$. At the same time, $v_1$ receives an updated information from $v_2$ on the receiver position. Ideally, if no processing and/or communication delay exists, vehicle $v_1$  always knows the exact location of $v_2$ at each time $t$. In practice, vehicle $v_2$ can send information on the signaling channel only with a period $\Delta T$. 
In this case, the deviations are then defined as
\begin{align}
\Delta x [t] = \delta x_2 [t]  - \delta x_2 [t - \Delta T] ,
\end{align} 
\begin{align}
\Delta y [t] = \delta y_2 [t] - \delta y_2 [t - \Delta T]  .
\end{align}
A short $\Delta T$ prevents the information to become outdated but it increases the signaling overhead, on the other hand a longer $\Delta T$ might be inefficient to track the dynamics but it reduces the overhead. We set  $\Delta T=20$ ms as the additional overhead of the signaling channel can be found in only 1.6 kb/s, assuming a 16 bit quantization of $x_2 [t]$ and $y_2 [t]$.

\section{Simulation Settings and Results }\label{sect:results}

\begin{figure}[!b]
	\centering 
	\includegraphics[width=0.7\linewidth, keepaspectratio]{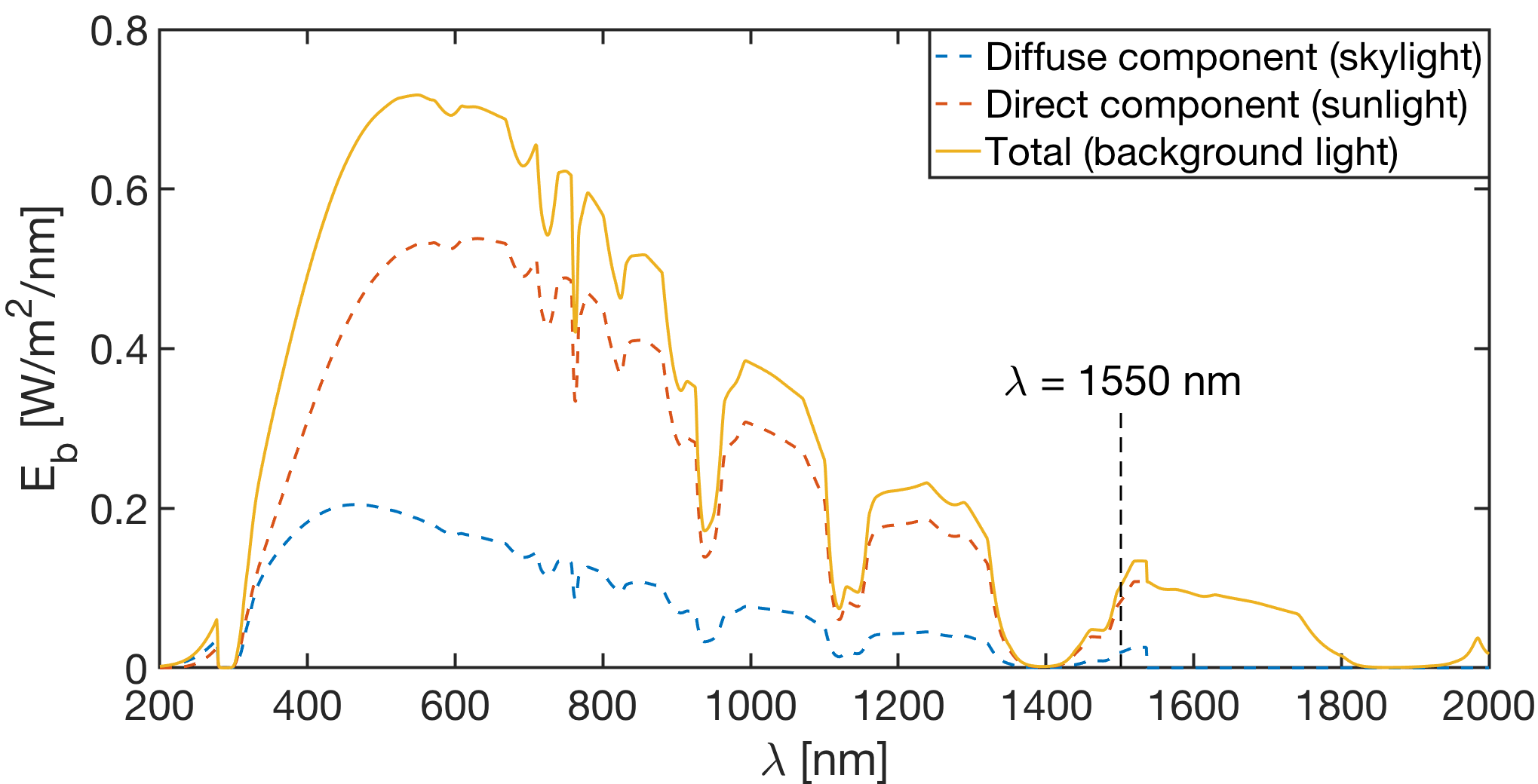}
	\caption{Direct, diffuse and total spectra of the solar radiation in the 200-2000 nm range, incident on a vertical surface positioned at 2 m above sea level (as the V2V FSO receivers), computed for the  Milan geographical zone, on July 20\textsuperscript{th}, 2018, 12 a.m., in clear sky conditions. Total irradiance: 475.7 $\mathrm{W/m^2}$. }
	\label{fig:solarradiation}
\end{figure}
We evaluate the performance of the proposed V2V FSO communication link in clear sky conditions, without any adverse weather effect, focusing only on the impact of vehicle misalignments and on the proposed solution for compensation. In the following, the reference communication bandwidth $B$ is set to 1 GHz as desirable for V2V scenarios. We assume to employ a laser transmitter at 1550 nm, for which there is a large availability of high-speed integrated Distributed Feed-Back (DFB) sources with emitting power ranging from 1 up to 10 mW (Class 1 lasers \cite{Hecht-2008}). As common, the laser spot diameter at its waist is in the order of few mm (2 mm in our case), which is compatible for integrated MEMS-based steering mirrors. The receiver comprises a GHz-bandwidth InGaAs PIN PD with an active area of 0.1 mm$^2$ covered by a focusing lens (or set of lenses) whose size is determined by the specific application. For V2V FSO, one of the design drivers is to reduce as much as possible the extra-size and weight of the system: in this regard, we make the hypothesis to use a receiving telescope of 1 mm$^2$ area, comparable with the one of a typical Gradient-Index (GRIN) lens. Given the receiver geometry under consideration, we evaluate the background light for a vertical surface, in the worst case possible, \textit{i.e.} when both the direct sunlight and the skylight are maximum and both impinge the receiver. Therefore, we obtain the solar spectrum for the geographical area of Milan, on July 20\textsuperscript{th}, 2018, at 12 a.m., assuming a very clear day (50 km visibility), depicted in Fig. \ref{fig:solarradiation}. 
We consider the receiver to be equipped with an optical filter with a bandwidth of 50 nm \cite{Macleod-2010}, centered around 1550 nm, which is in charge of reducing the background light.
The complete set of simulation parameters is reported in Table \ref{table:simulationparameters}. 
\begin{table}[!tb]
	\begin{center}
		\caption{Simulation parameters}
		\label{table:simulationparameters}
		\begin{tabular}{l | c  | c}
			\toprule 
			Parameter & Symbol & Value\\
			\midrule
			Emitted optical power & $P_0$ & 1, 10 mW \\
			Wavelength & $\lambda$ & 1550 nm \\
			Beam widths at the Tx & $W_{0,x}$,$W_{0,y}$ & 1 mm, 1 mm \\
			V2V distance & $z$ & 5:5:100 m \\
			Signal bandwidth & $B$ & 1 GHz\\
			Rx sensible area & $A_R$ & 1  mm$^2$  \\
			PD responsivity & $\eta$ & 0.8 A/W \\
			Optical filter bandwidth & $\Delta\lambda$ & 50 nm \\
			Solar irradiance & $I_{b}$ & 5.58 $\mathrm{W/m^2}$\\
			Noise Equivalent Power & $\mathrm{NEP}$ & 20 $\mathrm{pW/\sqrt{Hz}}$ \cite{Newport-highspeedrec-2013}\\
			\bottomrule
		\end{tabular}
	\end{center}
\end{table}

\begin{figure}[!tb]
	\centering
	\includegraphics[width=0.7\linewidth, keepaspectratio]{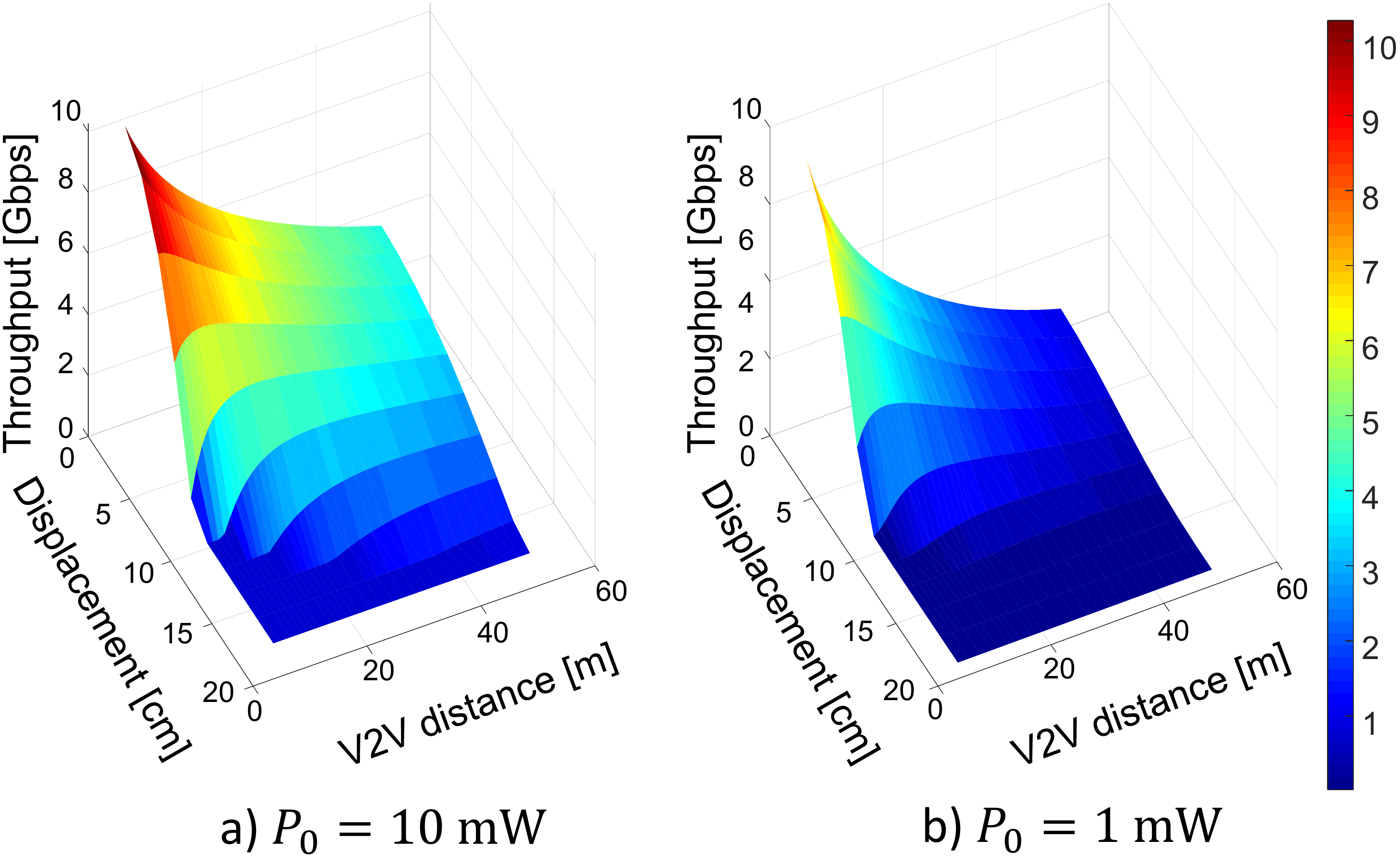}
	\caption{Impact of displacement on the throughput at different V2V distances for two transmitted power: a) $P_0=10$ mW, b) $P_0=1$ mW. }	
	\label{fig:displacement}
\end{figure}
\begin{figure}[!t]
	\centering
	\includegraphics[width=0.7\linewidth, keepaspectratio]{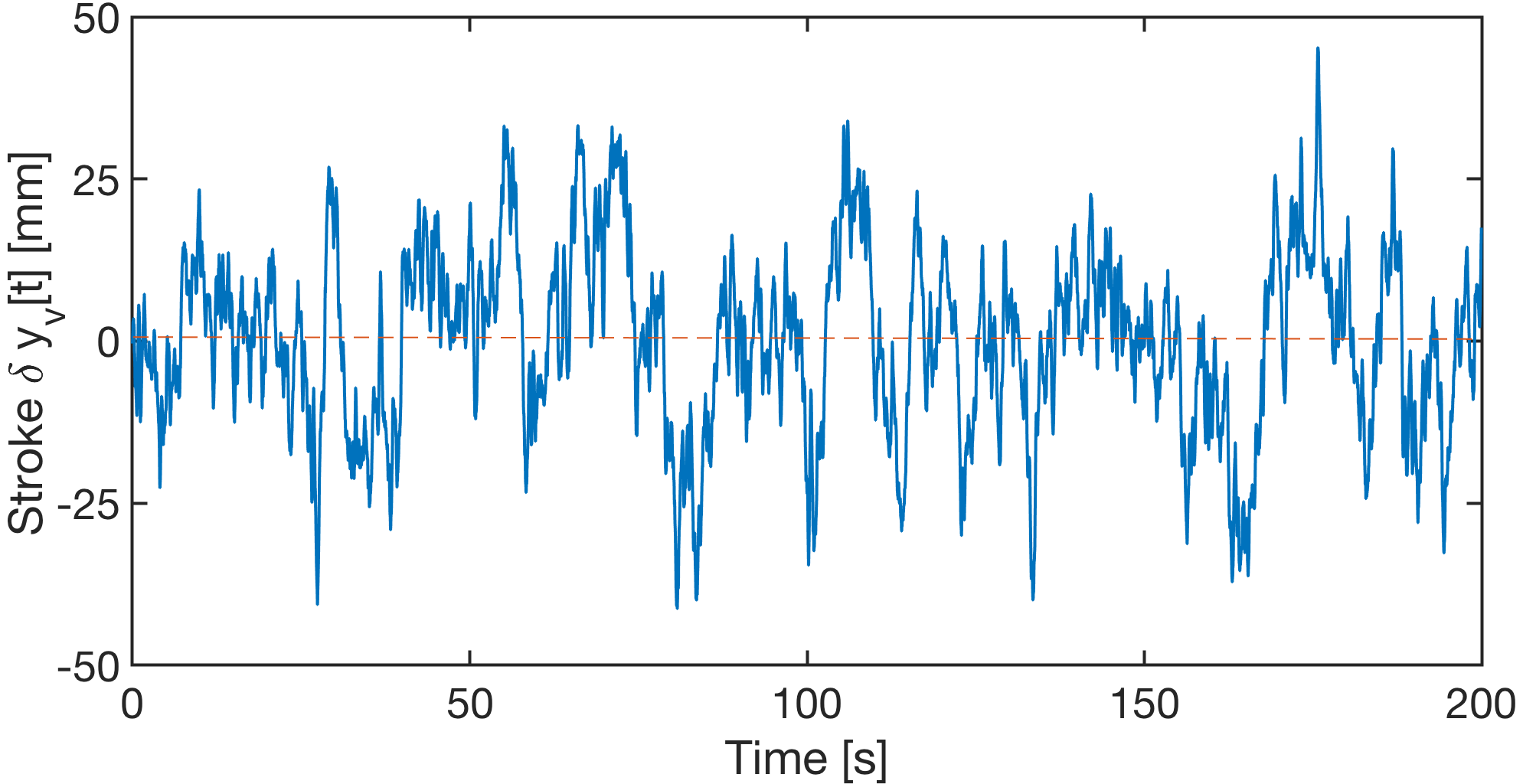}	
	\caption{Vehicle vertical perturbations $\delta y_v [t]$.}
	\label{fig:real_data}
\end{figure}
\begin{figure}[!t]
	\centering
	\includegraphics[width=0.7\linewidth, keepaspectratio]{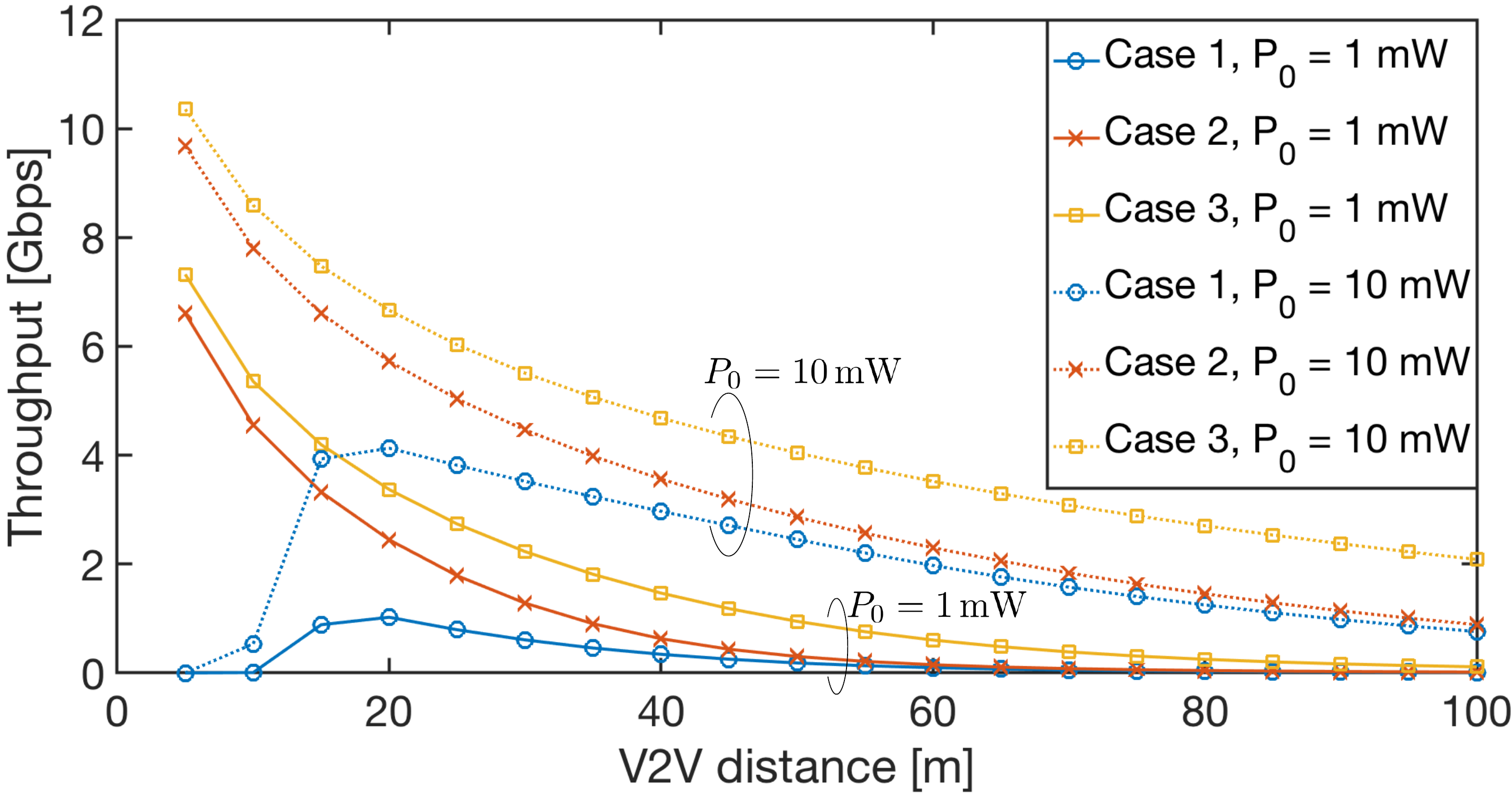}
\caption{V2V throughput for three different levels of compensation of vehicle dynamics and for two values of transmitted power $P_0$.}
	\label{fig:results}
\end{figure}

The first result we present concerns the scenario outlined in Sec \ref{subs:case1} (no compensation). The impact of misalignments on the channel throughput for different values of V2V distance is presented in Fig. \ref{fig:displacement}, for two values of transmitted power: $P_0=10$ mW (Fig. \ref{fig:displacement}.a) and $P_0=1$ mW (Fig. \ref{fig:displacement}.b). As it can be observed, a displacement of few centimeters causes a severe drop in the performance. For this reason, in mobility applications, a compensation mechanism is mandatory to orientate the laser towards the receiver within few centimeters of accuracy. 
The second result is the comparison of the achievable throughput of the V2V FSO link between two vehicles in a platoon formation, for the three case studies in Sec. \ref{sec:case studies}. We consider two vehicles aligned along the $x$-axis, \textit{i.e.}, $x_1 [t]= x_2 [t]$ $ \forall t$, and without any perturbation along this direction, \textit{i.e.}, $\delta x_1 [t]= \delta x_2 [t] = 0$ $ \forall t$.  On the other hand, the condition at rest along the $y$-axis for vehicle $v_1$ and $v_2$ are given by $\bar{y}_1 [t]= 1.4$ m and $\bar{y}_2 = 1.7$ m, respectively. Along this axis, the perturbations $\delta y_1 [t]$ and $\delta y_2 [t]$ hinder the pointing direction of the laser at $v_1$, as described in Sec. \ref{sec:geometrical settings}. These perturbations, which correspond to the strokes of the vehicle, are modeled as tenth order autoregressive processes AR(10), with filter coefficients calibrated on real measurements collected by a real vehicle (sedan). An example of perturbations $\delta y_v [t]$ is plotted in Fig. \ref{fig:real_data} for a duration of 200 seconds.
Fig. \ref{fig:results} provides the V2V FSO performance, for two reference values of transmitted power $P_0$ (1 and 10 mW) at different inter-vehicle distance, ranging from 5 to 100 m. Results in Fig. \ref{fig:results} highlight that a compensation mechanism is strictly required to fully exploit the V2V FSO channel, enabling the desired high-capacity (multi-Gbps)  communications when $P_0$ = 10 mW. A solution in which vehicles are able to compensate the nominal values of their condition at rest is feasible, but the unpredictable vibrations of vehicle dynamics still degrade the performance. On the other hand, a RF-assisted V2V FSO communication where vehicles periodically send information regarding their instantaneous dynamics is the most profitable and it can be implemented with a negligible signaling overhead. Note that the same simulation, not shown for lack of space, has been performed with negligible performance degradation for $\Delta T$ up 200 ms, relaxing the latency constraints on the RF link.

\section{Conclusions}

This paper proposes a Vehicle-to-Vehicle (V2V) communication scheme
based on Free-Space Optics (FSO) technology whereby the pointing of
lasers sources is based on the vehicles' information retrieved over
a low-rate Radio-Frequency (RF) link. In particular, as a first step,
this work has explored the impact of vehicle dynamics on laser pointing,
describing the effect of perturbations due to vehicle vibrations and
tilting. After showing that even a minor Tx-Rx
displacement (\textit{i.e.}, few centimeter)   is enough to cause a significant drop in the V2V throughput if not considered in the beam pointing, we
suggested a solution where the optical channel is assisted by a RF
control link to track and correct the pointing accordingly.
Numerical simulations confirmed that, by exploiting updated information on reciprocal Tx-Rx
locations obtained through the RF control link, the proposed RF-assisted FSO-based
V2V system allows for V2V communications at rates significantly higher then
those achieved by conventional techniques operating without the control
RF link, ranging from $\approx10$ Gbps (at a distance of 1 m) to $\approx2$ Gbps (at a distance of 50 m), thus providing a practical solution for V2V scenarios in high-mobility conditions. 
The latency analysis over both RF and FSO links, not covered here, requires further investigation. Nevertheless, while the latter is a very critical parameter for V2V FSO system design, the former can be significantly relaxed by considering lower update rate for signaling information.

Future research activities include also the analysis of the complete 3D V2V FSO communication system as well as the introduction of atmospheric effects which are known to severely affect the optical communication performance due to the energy absorption and scattering in the interaction with the atmospheric components.

\bibliographystyle{ieeetr}
\bibliography{Bibliography}
\end{document}